# Integrating GPS, GSM and Cellular Phone for Location Tracking and Monitoring

B. P. S. Sahoo and Satyajit Rath
*CSIR-Institute of Minerals & Materials Technology, Bhubaneswar, India*

*Abstract*— **The wide spread of mobiles as handheld devices leads to various innovative applications that makes use of their ever increasing presence in our daily life. One such application is location-tracking and monitoring. This paper proposes a prototype model for location tracking using Geographical Positioning System (GPS) and Global System for Mobile Communication (GSM) technology. The system displays the object moving path on the monitor and the same information can also be communicated to the user cell phone, on demand of the user by asking the specific information via SMS. This system is very useful for car theft situations, for adolescent drivers being watched and monitored by parents. The result shows that the object is being tracked with a minimal tracking error.**

## I. INTRODUCTION

Due to the ripeness in the wireless technology, location-tracking of objects and people in indoor or outdoor environments has received ample attention from researchers lately. There are various methods for identifying and tracking user position such as Cricket [10], Mote Track [7] or GPS [13]. GPS offers a scalable, efficient and cost effective location services that are available to the large public. However, the satellite emitted signals cannot be exploited indoors to effectively determine the location. The aim of this research is to track a user position in both indoor and outdoor environments with a minimal tracking error by incorporating a trans-receiver in the device which to be monitored. The tracking system can also be very useful for Intelligent Transportation System (ITS) [8, 12]. For example, it can be used in probing cars to measure real-time traffic data to identify the congesting area. It can also be a life saver in an emergency case to quickly and automatically report a vehicle position to a rescue agent when an accident happens to the vehicle. In addition, it can be attached to a vehicle with an anti-theft system to identify its location when it is stolen.

Our proposed system offers a real-time tracking system using a client-server approach. Our client can be any device, incorporated with the tracking antenna to identify device location information that is periodically transmitted to the base station via wireless medium [2, 12]. Our base station is a personal computer connected with a GPS receiver, GSM modem and interfacing software module written in C to filter the location information like altitude, longitude, speed, time and etc. and then converted into a format that can be displayed on PC, LCD and Mobile Phone as text message sent from GSM modem. This particular design has an advantage where the embedded system used for monitoring is not loaded with excess data because embedded application receives information only when there is a request from the user. This system has also another facility to monitor the path along which the device moves. This is optional, when the user wants, to see the path it receives the entire information from the GPS antenna otherwise it only receives the desired information. From application point of view, this environment tracking scheme can be utilized for various applications such as locating in-demand personnel like doctors or patients with vital sign sensor in hospital environment, Army, Air force, etc. [8]. It can also serve as a basis for context-aware application.

The remainder of the paper is organized as follows: In Section 2, we present primary aspects of technologies used in our designed system. In Section 3, the overall integrated system is described i.e. working methodology of our model. Finally, Section 4 summarizes the result of the project with future possible work we are going to attempt.

## II. TECHNOLOGY USED IN SYSETM DESIGN

Radio Frequencies based information and communication technologies, such as GPS, RFID tags and Bluetooth etc, have matured and become commercially available to potentially support automated data collection.

*A. GPS Receiver*
GPS is a Global Positioning System based on satellite technology. The activities on GPS were initiated by the US Department of Defense (DOD) in the early 1970s. The first of a nominal constellation of 24 GPS satellites was launched in

1978, and the system was fully operational in the mid- 1990s. Galileo is a European Union Global Positioning System, Glonass is a Russian and BeiDou is the Chinese system [5, 11]. GPS is comprised of three segments: space, control, and user equipment segments. The space segment consists of nominally 24 satellites that circulate the Earth every 12 sidereal hours on six orbital planes with four satellites in each plane that provide the ranging signals and data messages to the user equipment. The orbits are equally spaced 60° apart from each other, and each orbit has an inclination angle of 55°. GPS is a typical Medium Earth Orbit (MEO) system with an orbit altitude of approximately 20,200 km with regard to mean sea level. The control segment is responsible for monitoring and controlling the satellites subsystem health, status and signal integrity and maintains the orbital configuration of the satellites and consists of one Master Control Station and four unmanned monitor stations located strategically around the world. Finally, the user receiver equipment performs the navigation, timing, or other related functions (e.g., surveying) [1]. Each GPS satellite transmits signals at three distinct frequencies namely: L1 (1575.42 MHz), L2 (1227.60 MHz) and L5 (1176.45 MHz). The L1, L2 and L5 carrier frequencies are generated by multiplying the fundamental frequency by 154, 120 and 115, respectively. The measured transmitting times of the signals that travel from the satellites to the receivers are used to compute the pseudo ranges. Two levels of navigation and positioning are offered by the GPS. They are the Coarse Acquisition Code (C/A-code), sometimes called the "Civilian Code," and the Precise, or Protected Code (P-Code).The Course-Acquisition (C/A) code, sometimes called the Standard Positioning Service (SPS), and are a pseudorandom noise code that is modulated onto the L1 carrier. The precision (P) code, sometimes called the Precise Positioning Service (PPS), is modulated onto the L1, L2 and L5 carriers allowing for the removal of the effects of the ionosphere. Each satellite carries its own unique code string. The SPS is a positioning and timing service focusing on the civilian user and the PPS is a positioning, velocity, and timing service for military applications [12]. Each receiver has a clock, enabling it to measure the travel time of signals from the satellites. Measuring the signals TOF from the satellites, whose positions are known, enables the calculation of the receiver's 3-D position based on the trilateration principles. To calculate locations, the readings from at least four satellites are necessary, because there are four parameters to calculate: three location variables and the receiver's time [4]. To get metric or sub metric accuracy in positioning data (i.e. longitude, latitude, and altitude), a single GPS receiver is not sufficient; instead a pair of receivers perform measurements with common satellites and operate in a differential mode. DGPS provides sufficient accuracy for most outdoor tracking applications. In DGPS two receivers are used. One receiver measures the coordinates of a stationary point, called the base, whose position is perfectly known in the reference geodetic system used by GPS. The 3-D deviation between the measured and actual position of the base, which is roughly equal to the measurement error at a second receiver at an unknown point (called ''rover''), is used to correct the position computed by the latter [3]. The Real-Time Kinematic GPS (RTK GPS) can further enhance the positioning accuracy to centimeter (even millimeter) levels by combining the measurements of the signal carrier phases from both base and rover receivers with special algorithms.

*B. GSM Modem*

The idea of cell-based mobile radio systems appeared at Bell Laboratories in the early 1970s. In 1982 the Conference of European Posts and Telecommunications formed the Groupe Spécial Mobile (GSM) to develop a pan-European mobile cellular radio system (the acronym later became Global System for Mobile communications). In 1988, GSM system is validated [16]. The currently available technologies for mobile data transfer are Circuit Switched Data (CSD), High Speed Circuit Switched Data (HSCSD), General Packet Radio Systems (GPRS) and Third Generation (3G). Just as with audio transmission on landline phones, both CSD and HSCSD charges are based on the time spent using the dial-up connection. However, GPRS systems are sometimes referred to be as being always connected and dial-up modem connection is not necessary, so in comparison to CSD, immediacy is one of the advantages of GPRS (and SMS) [6]. GPRS is a non-voice value added service that allows data to be sent and received across a mobile telephone network that was designed to run on GSM, a worldwide standard for cellular communications. Data transmissions in the past were slow across the radio interfaces due to many propagation and reception problems. To create a broadband communications interface, GPRS was developed as a stepping-stone approach to other services like the Enhanced Data for a Global Environment (EDGE). GPRS is a packet switched "always on" technology supporting Internet Protocols (IP) with a theoretical maximum speed of up to 114kbps. Because GPRS uses the same protocols as the Internet, the networks can be seen as subsets of the
Internet, with the GPRS devices as hosts, potentially with their own IP addresses. In practice, connection speeds can be significantly lower than the theoretical maximum, depending upon the amount of traffic on the network and the type of handset being used [9]. Server unit uses GPS receiver to capture the current location and vehicle speed from the tracking antenna. Location and speed data provided by GPS is not in human understandable format. This raw data needs to be processed to convert it into useful information that can be displayed by monitor of the PC or in LCD display or it can be sent to the mobile phone through GSM modem. Software is required to process this raw data. GPS receiver can also provide information of altitude, time of GPS fix, status of GPS fix, and number of satellite used to compute current location information along with location and speed. GPS fix means last reported location. For tracking purpose only location and speed data is required for transmission. Other data provided by GPS receiver is used to determine the validity of location information.

*C. Tracking Software Module*
The raw data provided by the GPS receiver is captured by the software and processed to extract the required location and speed information. The microcontroller connected to the system is also responsible for monitoring the GPS receiver and GSM modem to receive and transmit the data to LCD and Mobile Phone as text message. This system holds all the required information that is to be transmitted to remote user (mobile phone) using AT commands, LCD and PC Monitor. It also controls data transmission module to exchange information with remote server. It actually acts as a bridge between GPS receiver, vehicle and remote server. It receives commands sent by remote server through data transmission/receiving module and performs corresponding action required by that remote server. The Location Tracking module is written and compiled using C language. The software performs three phases, reads message from mobile through GSM modem, the GPS position reading, and the GPS data formatted and transmitted to any display unit or return back a message giving location information via GPRS networks. The message reading phase prepares the module for reading and transmitting location information. It is composed of three functions. The baud rate used by the GPS (4800) and GSM (9600) are different. The first function reads the message from mobile through the GSM modem. The software reads the message from RS232 port and initializes signals for the microcontroller to switch the control to the GPS Receiver. After reading the location information from the GPS, microcontroller returns back its control from GPS to default GSM modem. Then it processed the raw data provided by the GPS receiver and extracted the required information and transmit it to the intended device.

*D. Subroutine- Send AT Command*
This subroutine is the basic routine which handles all the communication with RS232-GSM. All commands sent to module are sent using this subroutine. If the device responds with "OK", it means microcontroller can communicate with module. If device doesn't respond after expiration of timeout routine is restarted. If problem persists definitely something in hardware is damaged. After receiving "OK" response from module various parameters of module need to be initialized. SIM presence is checked by sending command "AT+CMGF=1" If device responds with "+CPIN: READY" message, SIM is ready to use. Any other response message will be considered as an error and routine will be restarted after expiration of timeout. When SIM card is ready, it is important to test whether module is connected to network or not. Network status can be tested with command "AT+CREG?" If module responds with "+CREG: 0, 1" module is connected to network and data can be sent over network. If any other response is received module keeps on checking for network status until it connects to network. Once it makes sure that module is connected to network, subroutine is terminated. Flow chart of the prototyped model and the subroutine sends AT command are shown in Fig. 1 and Fig. 2 respectively.

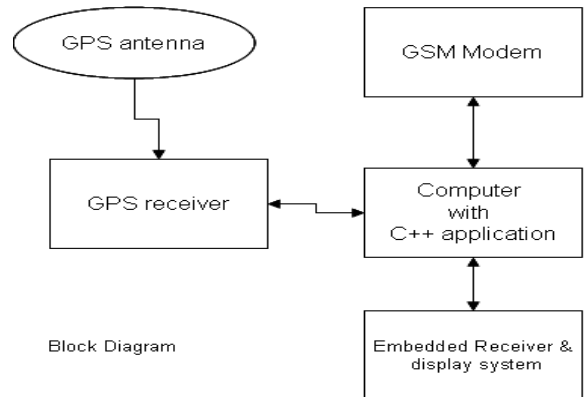

Figure 1. Block Diagram of Prototyped Model

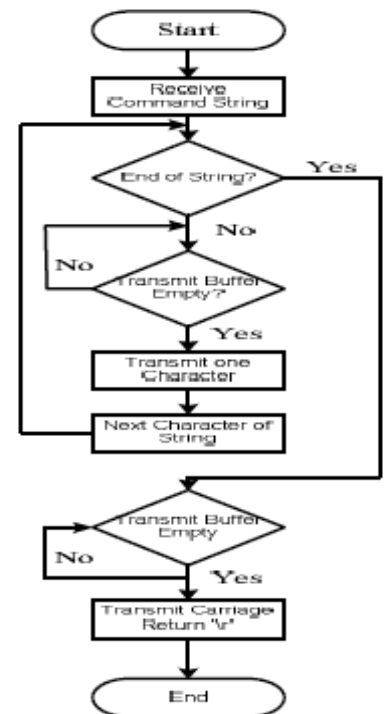

Figure 2. Flow Chart Subroutine sends AT command

III. WORKING METHODOLOGY
Overall system is partitioned into two major design units.
- Client Unit in Vehicle (Tracking Antenna)
- Server Unit connected to PC

The server part is the major part of the system and it will be installed into the PC. The sever part includes GPS receiver, GSM modem and one commodity hardware as shown in the Fig. 3.The client part will be installed into the vehicle or tracking device. The client part includes only one antenna as a trans-receiver. It is responsible for capturing information from satellite. The client unit is also responsible for transmitting this information wirelessly to Tracking Server located in the range. When the user wants to track the speed of vehicle, he/she needs to send a text message to the MSDN connected to the GSM modem by "SPEED". This is already defined in the server unit. Other than this message it can not recognize. Immediately after getting the message from the client the system starts read the current status of the vehicle and sends back the current speed to the mobile phone as a text message. In the same way if the user wants to know the location of the vehicle at any point, he/she needs to send a text message as "LOC". The same principle follows by the system to respond the user. Fig. 3 shows the screenshot of our implementation module. A case study on a particular track is displayed and is shown in Fig. 4. However, Fig. 5 shows altitude and longitude of one GPS Tracking modules. The same track can be viewed by using PC Monitor and Mobile Phone. Our system prototype has shown and tested on a trip in Bhubaneswar from Institute of Minerals and Material Technology to Utkal University.

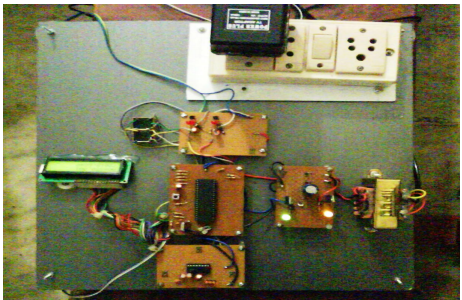

Figure 3. Implantation System Design

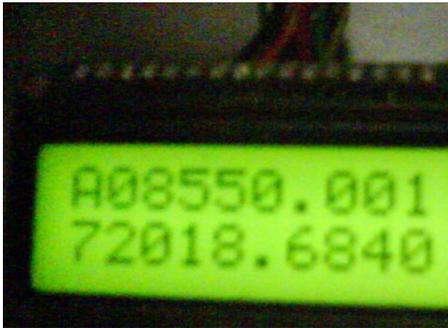

Figure 4. Location Tracked on LCD Screen

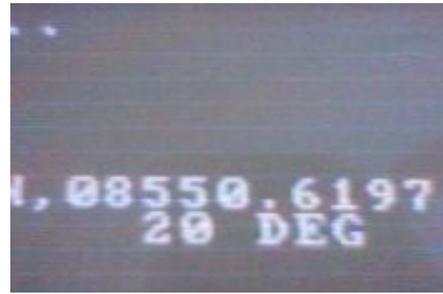

Figure 5. Tracked Location Displayed on PC Monitor

IV. CONCLUSION AND FUTURE DIRECTIONS

In this paper, we have proposed a location tracking system, using commodity hardware and software. Proposed system is an application framework of tracking management coupled with GPS and GSM technology for effective information delivery and management. The model permits real time control enabling corrective actions to be taken. The tracking system has shown the feasibility of using it for fleet management. It is believed that the application of this system will lead to important changes in the tracking management. The proposed system has numerous advantages. It can also be used for lost vehicle tracking when working with a car alarm system. In the future, we will plan to integrate other related devices in a vehicle such as sensors. The sensors report vehicle status information to our server, which can be useful for information processing and for intelligent tracking management.